\documentclass[epsf,twocolumn,showpacs,preprintnumbers]{revtex4}
\usepackage{graphics}
\usepackage{graphicx}
\usepackage{dcolumn} 
\usepackage{bm}
\usepackage{color}
\usepackage{epsfig}
\newcommand{\bco}{BaCrO$_3$}

\begin{document}
\title{Strain and Spin-Orbit Coupling Induced Orbital-Ordering in Mott Insulator BaCrO$_3$
}
\author{Hyo-Sun Jin$^1$, Kyo-Hoon Ahn$^1$, Myung-Chul Jung$^1$, and K.-W. Lee$^{1,2}$}
\affiliation{
 $^1$Department of Applied Physics, Graduate School, Korea University, Sejong 339-700, Korea\\
 $^2$Department of Display and Semiconductor Physics, Korea University, Sejong 339-700, Korea
}
\date{\today}
\pacs{71.20.Be, 71.30.+h, 71.20.Dg, 75.50.Ee }
\begin{abstract}
Using {\it ab initio} calculations, we have investigated an insulating tetragonally distorted perovskite
BaCrO$_3$ with a formal $3d^2$ configuration, 
the volume of which is apparently substantially enhanced by a strain 
due to SrTiO$_3$ substrate.
Inclusion of both correlation and spin-orbit coupling (SOC) effects leads to
a metal-insulator transition and in-plane zigzag orbital-ordering (OO) of 
alternating singly filled $d_{xz}+id_{yz}$ and $d_{xz}-id_{yz}$ orbitals, 
which results in a large orbital moment $M_L\approx-0.78 \mu_B$ antialigned to the spin moment 
$M_S\approx2|M_L|$ in Cr ions. 
Remarkably, this ordering also induces a considerable $M_L$ for apical oxygens. 
Our findings show metal-insulator and OO transitions, 
driven by an interplay among strain, correlation, and SOC, 
which is uncommon in $3d$ systems.
\end{abstract}
\maketitle

\section{Introduction}
In condensed matter physics the interplay among strain, strong correlation, and
spin-orbit coupling (SOC) has been becoming extremely important.\cite{mann,hwang12,will}
This leads to abundant unconventional phenomena such as magnetic ordering, 
interface superconductivity, metal-insulator transitions, and more recently
topological insulators.\cite{GdN,chal,han,wang,tan,reich,jyu13} 
In a Mott transition, in particular, the ratio $U/W$ of correlation strength $U$ 
to bandwidth $W$ determines charge-, spin-, and orbital-orderings.
$W$ can be controlled by external factors such as doping, pressure,
and strain by a proper choice of substrate.
The strain has been of great interest for applications of band engineering.\cite{GdN,chal,han,wang,tan}
Recently, another type of the Mott insulator induced by correlation effects 
which are assisted by spin-orbit coupling (SOC), named relativistic Mott insulator, 
has been observed in cubic double perovskite Ba$_2$NaOsO$_6$ and a few 
Ir-based oxides, which show substantially strong SOC.\cite{LP07,yu}
The relativistic Mott transition was also proposed to occur even in the $4d$ Li$_2$RhO$_3$ system.\cite{luo}
In this transition, SOC leads to the removal of degeneracy in partially filled bands 
or narrowing of the bandwidth, and then strong correlation selects a filled orbital.
However, pure correlation effects often lead to a metal-insulator transition
in $3d$ systems, where the strength of SOC is negligible compared to that of correlation.
Here, we will address a strain and SOC-driven Mott transition, 
accompanied by orbital-ordering (OO), in the $3d^2$ BaCrO$_3$ thin film 
synthesized recently.\cite{bco_exp}
In this system, strength of SOC is indeed tiny, but crucial to open a gap.
This has been rarely observed in $3d$ systems.

Tetravalent $d^2$ chromate perovskites have recently drawn attention due to their atypical 
and controversial properties.\cite{cco_exp,cco_the,sco_exp,sco_the,srcro3}
Although the orthorhombic antiferromagnet CaCrO$_3$ is metallic, 
it was suggested that correlation effects play an important role.\cite{cco_exp,cco_the}
Another metallic antiferromagnet SrCrO$_3$ has a tetragonally distorted structure
due to partial OO of $(d_{xy})^1(d_{xz}d_{yz})^1$.\cite{sco_exp,sco_the}
Through correlated first principles calculations, 
Gupta {\it et al.} proposed that lattice distortions in an ultrathin SrCrO$_3$ film 
lead to OO, which in turn induces ferroelectricity.\cite{srcro3}
Recently, Zhu {\it et al.} synthesized the missing member, \bco, 
of the tetravalent chromate perovskites, on SrTiO$_3$ (001) surfaces.\cite{bco_exp}
As observed in the isostructural SrCrO$_3$,\cite{sco_exp} the distortion factor of $a/c$ 
is as small as approximately 0.5 \%, 
and this prevents the CrO$_6$ octahedron from rotating and tilting (see below).
The magnetic measurements show hysteresis with a small coercive field and small saturated
moment of 0.028 $\mu_B$ with Curie temperature $T_C=25$ K, 
indicating a weak ferromagnet (FM) or canted antiferromagnet (AFM).
The resistivity data follow an insulating behavior with a kink at $T\approx200$ K
and indicate an activation energy gap of $\sim$ 0.38 eV.
They also performed correlated first principles calculations using 
the local spin density approximation plus Hubbard $U$ (LSDA+U) approach,
but failed to obtain the observed insulating state.\cite{bco_exp} 
Motivated by the failure of LSDA+U, Giovannetti {\it et al.} 
carried out dynamical mean field calculations
and concluded that a Jahn-Teller distortion in the plane of the tetragonal cell with
two different Cr-O bond lengths would lead to an insulating state.\cite{bco_dmft}

The nature of the insulating state of \bco~ can be unveiled from first principles
calculations, including $U$ and (relativistic) SOC effects,
using an accurate all-electron full-potential electronic method of {\sc wien2k}.\cite{wien2k}
Considering only $U$ and SOC simultaneously throughout the LSDA+SOC+U approach, 
we obtained the experimentally observed insulating state,\cite{bco_exp}
indicating a crucial role of SOC.
In contrast to the Jahn-Teller distortion incorporated scenario,\cite{bco_dmft}
our results show that this insulating state
is induced by the small structure distortion and SOC activated by narrowing $W$, 
resulting from the enlarged volume due to the strain of the substrate.
The accompanying OO stabilizes the C-type AFM (C-AFM), 
which is antiparallel in-plane and parallel along the $c$-axis,
and leads to an unusually large orbital moment $M_L$ equal to half of the spin moment $M_S$ 
in this system. 
Interestingly, as shown below, this feature also drives 
a considerable $M_L$ in the apical oxygens. 

\begin{figure}[tbp]
\vskip 4mm
{\resizebox{8cm}{6cm}{\includegraphics{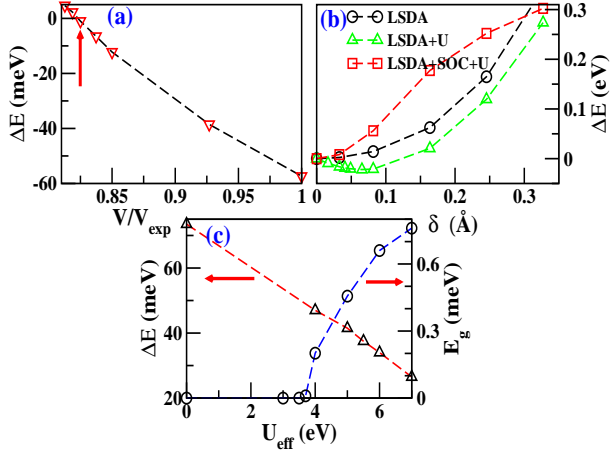}}}
\caption{(Color online) (a) Difference in energy between the cubic and the tetragonal phases,
as a function of volume. The arrow denotes the optimized volume.  
(b) Change in energy versus  a variation of the in-plane Jahn-Teller distortion $\delta$. 
Here $\delta$ is the difference between short and long Cr-O bond lengths.
The results of LSDA+SOC, very similar to those of LSDA, are not shown here.
The Jahn-Teller distortion is only stable in the LSDA+U calculations.
(c) Energy difference between FM and C-AFM (left side), and the energy gap (right side)
 as a function of the effective Coulomb repulsion $U_{eff}$ in LSDA+SOC+U.
}
\label{str}
\end{figure}

\section{Crystal structure and calculation method}
According to the experiment by Zhu {\it et al.}, the tetragonal perovskite phase
appears to be unstable in the bulk.\cite{bco_exp} 
We optimized the crystal structure to investigate stability of the tetragonal structure.
During the optimization in the tetragonal phase, 
the ratio $a/c$ was fixed at the value obtained in the experiment.\cite{bco_exp}
The structural parameters were optimized until forces were smaller than
2 mRy/a.u. 
As shown in Fig. \ref{str}(a), our optimized volume is
about 18\% less than that of the experiment.
Considering that the L(S)DA usually underestimates an experiment volume 
by at most several percents, as observed for the similar compound SrCrO$_3$,\cite{sco_the}
this difference is significant.
This implies that the SrTiO$_3$ substrate significantly expands the volume
of \bco, consistent with the instability of the bulk state.\cite{bco_exp}
Remarkably, near the optimized volume the cubic and tetragonal phases are nearly degenerate.
We will return to this issue below.
When $c$-axis lattice parameter in perovskites decreases, 
the in-plane lattice parameter tends to increase to prevent the volume from varying.\cite{wep96} 
This leads to the distortion from the cubic to the tetragonal structures.
For the experiment volume, the tetragonal phase has an energy 57 meV lower than
the cubic phase, consistent with the experiment.
We also investigated a possible tilting and rotating of the CrO$_6$ octahedron.
However, our trials to obtain this gadolinium orthoferrite GdFeO$_3$-type structure 
always converged to the tetragonal phase observed in the experiment,\cite{bco_exp}
indicating that the GdFeO$_3$-type distortion is not energetically favored in this system.
From hereon, unless stated otherwise, we use the experiment lattice parameters 
$a=4.09$ \AA~ and $c=4.07$ \AA.\cite{bco_exp}

In {\sc wien2k}, the basis size was determined by R$_{mt}$K$_{max}$=7
and augmented-plane-wave sphere radii (2.5 for Ba, 2.01 for Cr, and 1.82 for O).
The Brillouin zone was sampled with up to a dense $k$-mesh
of $11\times 11\times 15$.
Additionally, linear response full-phonon calculations were performed
with {\sc quantum espresso}.\cite{qe}
These calculations were carried out with a 2$\times$2$\times$2 $q$-mesh,
a 14$\times$14$\times$14 $k$-mesh, an energy cutoff of 40 Ry,
and a charge cutoff of 400 Ry.

\begin{figure}[tbp]
{\resizebox{8cm}{6cm}{\includegraphics{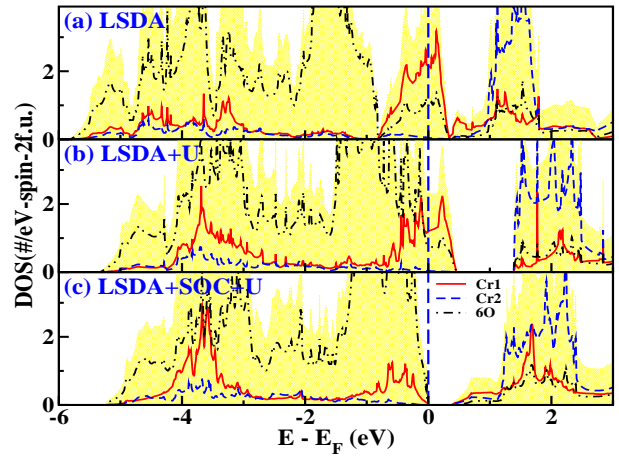}}}
\caption{(Color online) Atom-resolved densities of states (DOSs)
for the spin-up channel in LSDA, LSDA+U, and LSDA+SOC+U 
(from top to bottom) in the C-AFM,
containing two formula units. 
For LSDA+U and LSDA+U+SOC, $U_{eff}=5$ eV was used.
The shaded region denotes the total DOS.
In LSDA (and LSDA+SOC), the $t_{2g}$-$e_g$ crystal field splitting is $\sim$1.5 eV,
identical to the exchange splitting of the $t_{2g}$ manifold.
(The LSDA+SOC DOS, which is nearly identical to that of LSDA, is not shown here.)
}
\label{dos}
\end{figure}

\begin{figure}[tbp]
{\resizebox{8cm}{6cm}{\includegraphics{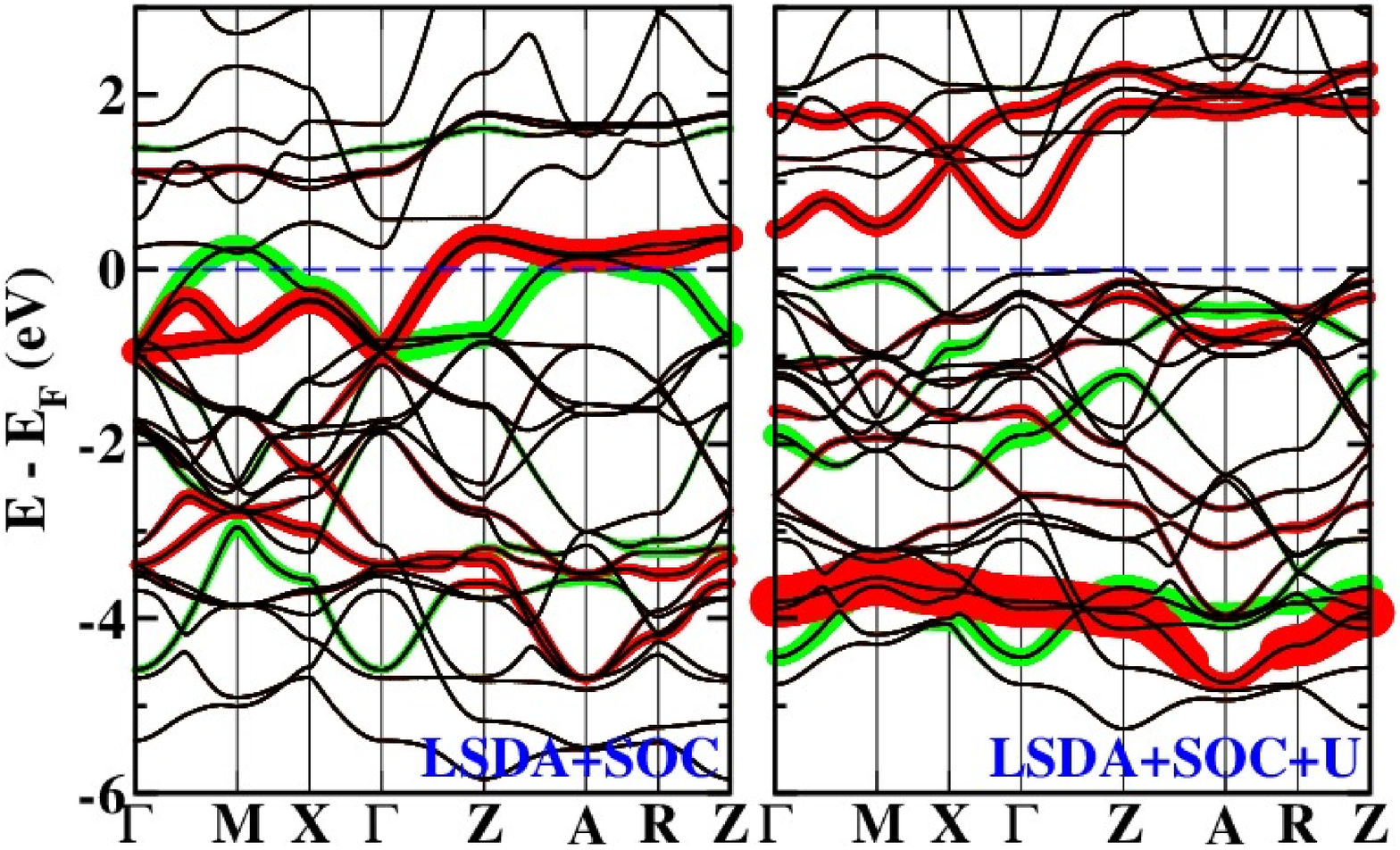}}}
\caption{(Color online) C-AFM band structures of LSDA+SOC (left) and LSDA+SOC+U 
at $U_{eff}=5$ eV (right), 
plotted in the basal plane. The highlighted fatbands indicate
the $d_{xy}$ (green color) and $d_{\pm1}$ (red color) characters of the spin-up Cr.
As expected from the weak SOC strength, the LSDA+SOC band structure is nearly identical
to that of LSDA (not shown here).
In the 45$^\circ$ rotated supercell, the $X(A)$ and $M(R)$ points are 
($\frac{\pi}{a}$,$\frac{\pi}{a}$,$\xi$) and ($\frac{\pi}{a}$,0,$\xi$), respectively.
$\xi$ is zero for the first symbols and $\frac{\pi}{c}$ for the symbols
in parentheses.
}
\label{band}
\end{figure}

\begin{table}[bt]
\caption{ Individual spin $M_S$, orbital $M_L$, and net $M_{net}$ moments 
 of the C-AFM (in units of $\mu_B$), 
obtained from LSDA+SOC and LSDA+SOC+U at $U_{eff}=5$ eV calculations.
The moments of the planar oxygens are zero by symmetry. 
}
\begin{center}
\begin{tabular}{cccccccc}\hline\hline
  ion    &  \multicolumn{3}{c}{LSDA+SOC}& ~  &\multicolumn{3}{c}{LSDA+SOC+U}  \\
                         \cline{2-4}\cline{6-8}
     ~   &  $M_{S}$ & $M_{L}$ & $M_{net}$ & ~&$M_{S}$ & $M_{L}$ & $M_{net}$ \\\hline
  Cr     & 1.677 & --0.040 & 1.637  & ~& 1.451 & --0.781 & 0.670  \\
apical O & 0.073 & 0.004  & 0.077 & ~& 0.095 & --0.075  & 0.020 \\\hline
\end{tabular}
\end{center}
\label{table1}
\end{table}

\section{Results}
The results of the generalized gradient approximation plus Hubbard $U$ (GGA+U), 
performed by Giovannetti {\it et al.},\cite{bco_dmft} show that the in-plane Jahn-Teller
distortion is stable. This is consistent with our LSDA+U calculations, 
as shown in Fig. \ref{str}(b).
However, the addition of SOC to the LSDA+U approach
relieves the Jahn-Teller distortion in both metallic and insulating states.
This rules out the scenario of Mott transition driven by the Jahn-Teller distortion
in this system, which was proposed by Giovannetti {\it et al}.

In the CrO$_6$ octahedron, 
the tetragonal distortion, shortening of $c$ lattice parameter,
lowers the center of $d_{xy}$ and results in the splitting of the $t_{2g}$ manifold 
into the singlet $d_{xy}$ and doublet $d_{xz}/d_{yz}$.
Thus, for this $d^2$ configuration, the doublet is half-filled 
in a spin ordered state.

Initially, we applied the on-site Coulomb repulsion $U$ to the Cr ions 
to obtain the observed insulating state, using the LSDA+U approach,
in which the effective on-site Coulomb repulsion $U_{eff}=U-J$ was used
where $J$ is the Hund's exchange integral.
As observed in the previous calculation,\cite{bco_exp} inclusion of only $U_{eff}$ 
cannot produce an insulating state even for large values such as $U_{eff}=8$ eV, 
which is already beyond a reasonable value for this system.
Therefore, we add to the LSDA+U approach the effects of SOC 
in the magnetization direction along the $c$-axis, which is the high symmetry
direction in the crystal structure.
In the $t_{2g}$ manifold, SOC transforms \{$d_{xy}$, $d_{yz}$, $d_{xz}$\} 
into \{$d_{xy}$, $d_{\pm1}=d_{xz}\pm id_{yz}$\}.\cite{LP07} 
In the LSDA+SOC+U calculations, an energy gap appears 
at the critical value $U_{eff}^c\approx3.7$ eV,\cite{dc}
and finally reaches to the experimentally observed gap of $\sim0.4$ eV at $U_{eff}=5$ eV.
Figure \ref{str}(c) shows the change in the energy gap as a function of $U_{eff}$ 
in the C-AFM. 
The $U_{eff}^c$ is somewhat larger than the value obtained from the constrained
random-phase approximation,\cite{bco_dmft}
since $U$ acts on the mixture of Cr $3d$ and O $2p$ due to
the strong $p-d$ hybridization, which can be measured in the DOS of Fig. \ref{dos}. 
It is noteworthy that there is a regime around $U_{eff}^c$ 
in which both metallic and insulating solutions are obtained, 
as is extensively discussed in literature.\cite{LP05}
However, for this system an insulating state is always energetically favored 
over a metallic state at the same $U_{eff}$ in this regime.

To disclose the nature of the insulating state, we analyze the electronic
and magnetic structures in detail. 
Figure \ref{str}(c) shows that the C-AFM is energetically favored
over FM, which has a much lower energy than
the nonmagnetic state (by 216 meV/f.u. in LSDA+SOC), 
regardless of the strength of $U_{eff}$. 
As $U_{eff}$ increases, the difference in energy ($\Delta E$) per formula unit 
between these two states 
monotonically decreases with a slope of $\Delta E/U_{eff}=-6.6\times 10^{-3}$.
Using a simple Heisenberg model $H=-J\sum_{i,j}\vec{S_i}\cdot\vec{S_j}$,
for this spin $S=1$ configuration the superexchange parameter $J$
is estimated by $J=\Delta E\approx 42$ meV at $U_{eff}=5$ eV.
The C-AFM alignment has been observed in both CaCrO$_3$ and SrCrO$_3$.\cite{cco_exp,sco_exp}
 
\begin{figure}[tbp]
\vskip 8mm
{\resizebox{7cm}{6cm}{\includegraphics{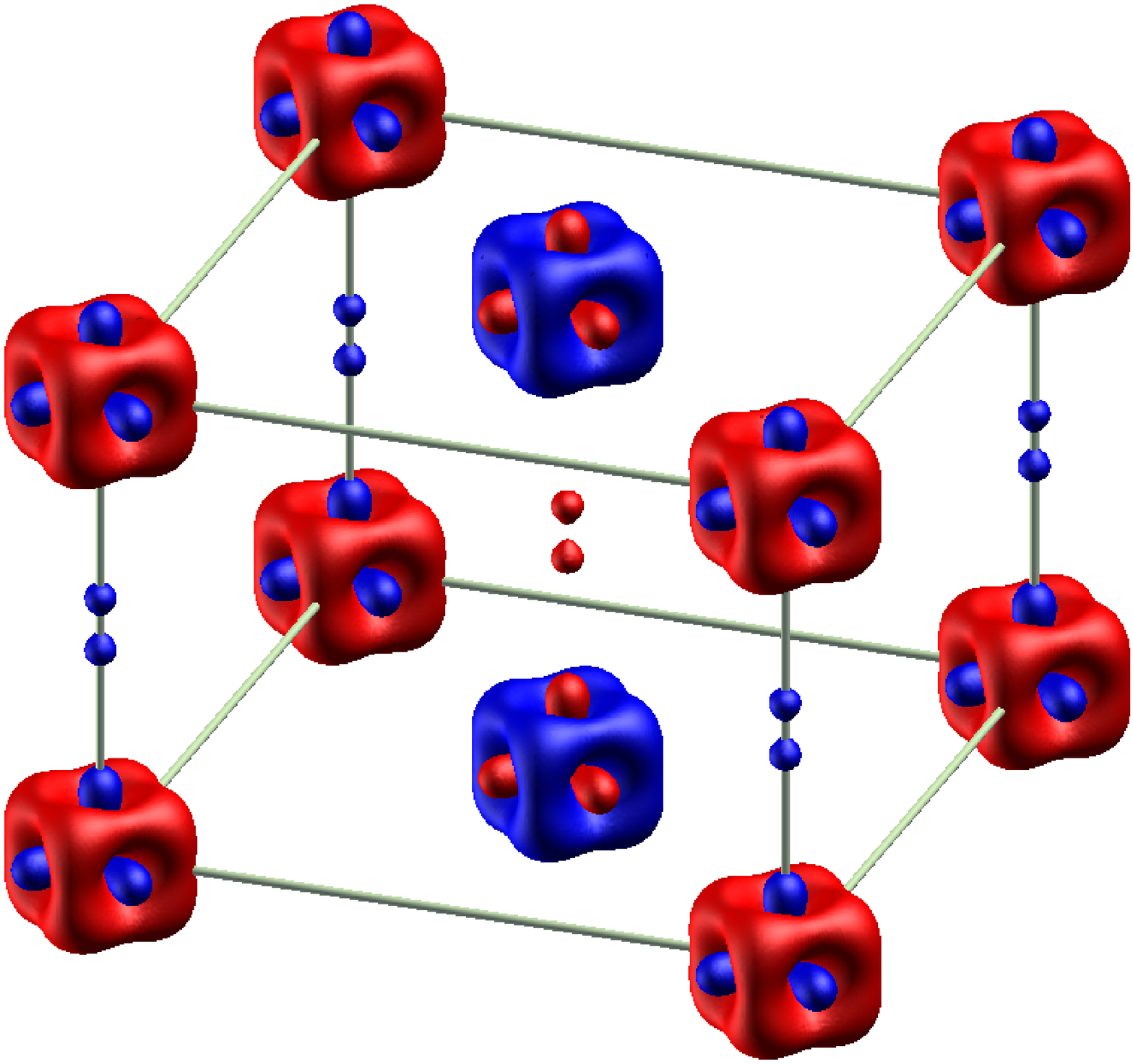}}}
{\resizebox{7cm}{6cm}{\includegraphics{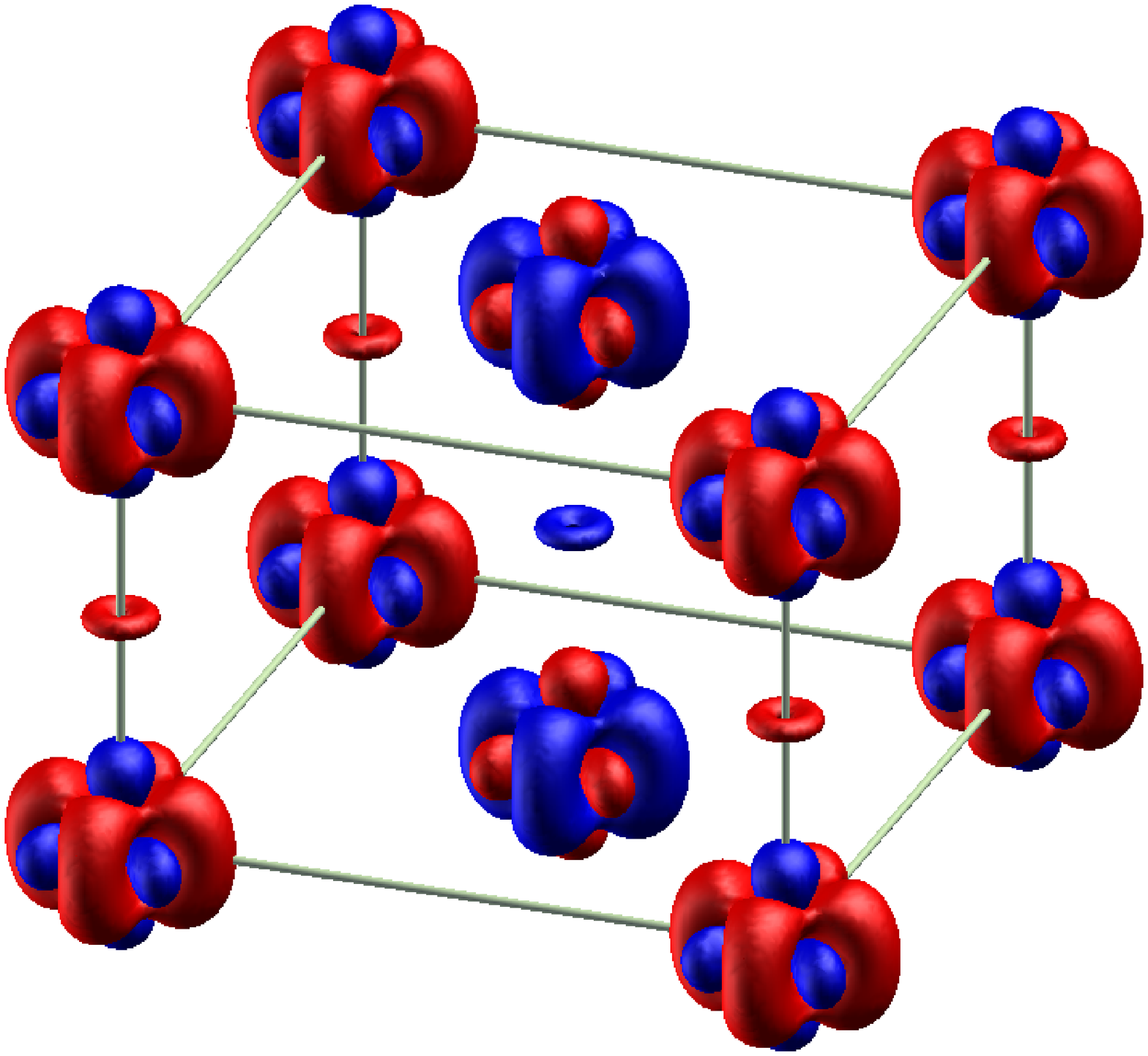}}}
{\resizebox{5.5cm}{3.0cm}{\includegraphics{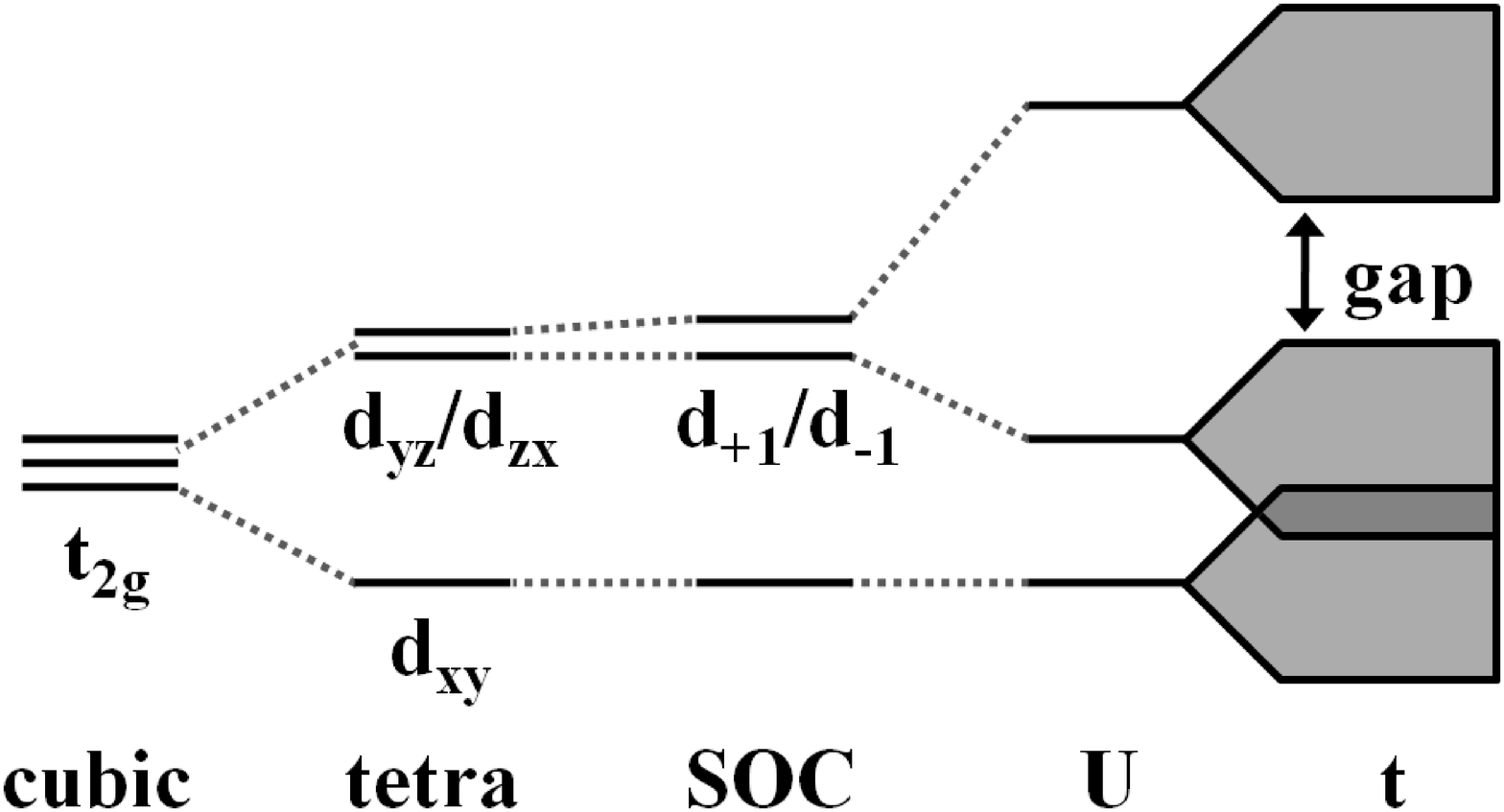}}} 
\caption{(Color online) Spin density plots of Cr ions
for LSDA+SOC (top) and LSDA+SOC+U (middle) 
in the $\sqrt{2}\times\sqrt{2}$ supercell.
The isosurface of LSDA+SOC+U is at 0.063$e/$\AA$^3$, while
a slightly larger isovalue of 0.1$e/$\AA$^3$ is used  in LSDA+SOC
for a better visualization.
The spin density of the apical oxygens is clearly visible, while
that of the planar oxygens is invisible for this isovalue.
The different colors represent the opposite spin orientation
in the C-AFM state.
Bottom: Schematic evolution of the spin-up $t_{2g}$ manifold in the cubic symmetry,
through tetragonal distortion, SOC, Coulomb repulsion, and dispersion
due to hopping $t$.
}
\label{spin}
\end{figure}

Now, we focus on the C-AFM alignment in both the metallic
and insulating phases, where the latter is obtained from the LSDA+SOC+U
calculations at $U_{eff}=5$ eV.
The total and atom-projected DOSs for both the metallic and insulating phases
are displayed in Fig. \ref{dos}.
The corresponding band structures with the emphasized fatbands of
the $d_{xy}$ and the $d_{\pm1}$ orbitals are given in Fig. \ref{band}.
In fact, the electronic structures of LSDA+SOC are nearly identical
to those of LSDA (not shown here), since the strength of SOC is only several meV
near the Fermi energy.
In the metallic state, the partially filled $t_{2g}$ manifold is separated from 
the unfilled $e_{g}$ manifold in most regimes, 
but the $d_{xy}$ band touches the bottom of the conduction band
with a $d_{z^2}$ character at the $M$ point.
As shown in the DOS, $W$ of the $\frac{2}{3}$-filled $t_{2g}$ manifold 
consistent with the $d^2$ configuration, is about 1.3 eV.
This leads to an estimated value of the nearest neighbor hopping parameter 
$t\approx0.1$ eV.
This value is a half of that for the nonmagnetic state, indicating a considerable
narrowing due to the AFM alignment.
Our calculated moments are listed in Table \ref{table1}.
The spin moment $M_S$ for Cr is 1.68 $\mu_B$,
substantially lower than the formal value of 2 $\mu_B$ 
due to the strong $p-d$ hybridization.
These features can be visualized in a spin-density isocountour plot, as displayed
in the top panel of Fig. \ref{spin}.
In the metallic phase, the minority of Cr atoms has a $d_{z^2}+id_{x^2-y^2}$ character,
while the majority of Cr indicates an equally occupied $t_{2g}$ manifold.
This clear visibility of the minority density indicates a strong $pd\sigma$ hybridization.

In the insulating state, the upper Hubbard band is at the bottom of the conduction 
band, while the occupied $d_{xy}$ and the lower Hubbard band 
are mostly around --4 eV, as shown in the right panel of Fig. \ref{band}.
The corresponding spin-density plot is displayed in the middle panel of Fig. \ref{spin}.
The Cr majority character represents a mixture of $d_{xy}+id_{\pm1}$.
As given in Table \ref{table1}, the Cr orbital moment $M_L\approx-0.78$ $\mu_B$ 
is antialigned to the corresponding $M_S$, leading to the net moment $M_{net}=0.67$ $\mu_B$.
This antialignment indicates that the lower Hubbard band has mostly the $d_{-1}$ character 
for the spin-up Cr, but the $d_{+1}$ character for the spin-down Cr.
This zigzag OO in the $ab$-plane is consistent with the C-AFM alignment.
A similar OO was observed in the spinel ZnV$_2$O$_4$, as including both $U$ and SOC.\cite{tcher,mait}
Although some $3d$ systems show an unquenched $M_L$,\cite{min,huang02,huang04,hwu,fecrs,mnvo,bart}
this substantially large $M_L$ is rare in $3d$ systems.
Interestingly, the spin density of the apical oxygens clearly indicates
the $p_x\pm ip_y$ character, even though the $M_S$ of the apical oxygens is only 0.095 $\mu_B$.
$M_S$ of the apical oxygen is nearly compensated by the considerable 
$M_L=-0.075$ $\mu_B$, which is unprecedented for an oxygen ion.\cite{huang02}
However, recently this unusual feature has been observed in a $5d$ system.\cite{LP_OO}

\begin{figure}[tbp]
\vskip 8mm
{\resizebox{8cm}{6cm}{\includegraphics{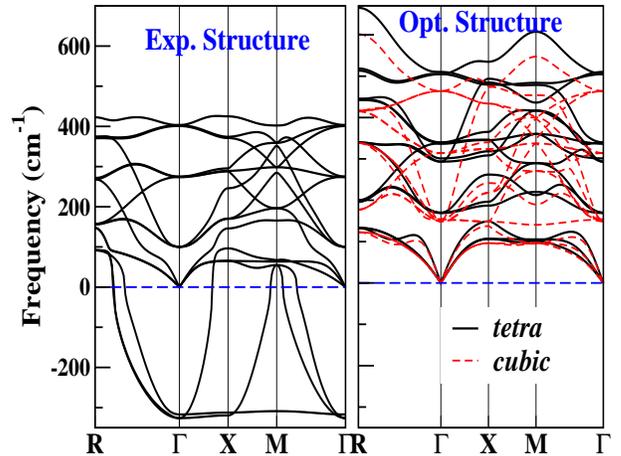}}}
\caption{(Color online) GGA phonon spectra with the experimental (left) 
and our optimized volumes (right).
The experimental spectrum shows imaginary frequencies throughout all regimes,
indicating a significant instability in structure. 
}
\label{phonon}
\end{figure}

\section{Discussion and Summary}
We also performed LSDA+SOC+U calculations for our optimized volume.
As expected from widening $W$ of the $t_{2g}$ manifold by $\sim25$\%  
due to compression of the volume, 
in the optimized structure a gap appears at $U_{eff}^c\approx5.5$ eV, about
50 \% larger than in the experiment structure. 
Thus, if the optimized volume can be attained experimentally, this system would be metallic. 
These results suggest that change in strain (or applying a pressure) 
leads to an insulator-metal transition in this system.
  
To investigate the stability of these structures, we carried out linear response 
full-phonon calculations
for both the experiment volume and our optimized volume.
The results are displayed in Fig. \ref{phonon}.
In good agreement with the instability in the bulk,\cite{bco_exp} 
there are several imaginary frequencies in the experiment volume.
However, our results show that in the optimized volume 
both the tetragonal and cubic perovskites are stable.
This suggests that these phases would be achieved by a proper choice 
of substrate or a high pressure technique, requiring further experiments.
Considering the tiny difference in energy ($\le$1 meV) between the cubic and 
the tetragonal phases, 
our results imply that the cubic structure would be stabilized by quantum fluctuations.\cite{vand} 

In summary, we have carried out {\it ab initio} calculations including both correlation 
and SOC effects to investigate 
a tetragonally distorted perovskite BaCrO$_3$, 
whose the volume seems to be considerably enhanced by the strain of the SrTiO$_3$ substrate.
Inclusion of both SOC and $U$ causes this $d^2$ system to change to insulating 
at $U_{eff}^c\approx 3.7$ eV,
though the metallic state remains unchanged on applying $U$ only. 
This indicates a crucial role of SOC in this Mott insulator, 
which is uncommon in $3d$ systems.
In the insulating state, one electron occupies $d_{xy}$ and the other alternatively occupies 
$d_{xz}+id_{yz}$ or $d_{xz}-id_{yz}$ orbital, 
resulting in an unquenched Cr orbital moment $M_L$ which is antialigned 
to the Cr spin moment $M_S\approx2|M_L|$.
This is consistent with the ground state of C-AFM, 
implying that the observed small moment is due to a little canted spin. 
Remarkably, due to the spin and orbital orderings of the Cr ions 
the spin density of the apical oxygen ions clearly shows the admixture of $p_x\pm ip_y$ characters, 
leading to the considerable spin and orbital moments antialigned each other.
These may be directly probed by further experiments like x-ray resonant spectroscopy.\\

\section{Acknowledgments}
We acknowledged C. Kim for useful discussions on a thin film growth,
and W. E. Pickett and D. Kashinathan for useful communications on an OO,
and HB Rhee for a critical reading for the manuscript.
This research was supported by NRF-2013R1A1A2A10008946.


\begin{thebibliography}{10}
\bibitem{dc} As including $U$, we used two popular double counting schemes
of the so-called around mean field (AMF)\cite{amf} and the fully-localized limit (FLL)\cite{fll}.
However, in the LSDA+SOC+U calculations, only the AMF scheme leads to
open a gap above $U_{eff}^c$. This is consistent with the fact that AMF is often
more suitable than FLL for the moderately correlated systems,
as discussed in detail in literature.\cite{LP05}
Thus, we will address the results from AMF, here.
\end{thebibliography}
\end{document}